\documentclass[a4paper,10pt,twoside]{cpc-hepnp}

\usepackage{multicol}
\usepackage{graphicx}
\usepackage{amssymb,bm,mathrsfs,bbm,amscd}
\usepackage[tbtags]{amsmath}
\usepackage{lastpage}
\usepackage{bm}

\newcommand {\ihep} {1}
\newcommand {\gscas} {2}
\newcommand {\tshu} {3}
\newcommand {\nju} {4}
\newcommand {\zzu} {5}
\newcommand {\pku} {6}
\newcommand {\lnu} {7}
\newcommand {\gxnu} {8}
\newcommand {\ustc}{9}
\newcommand {\sdu} {10}

\newcommand{\etal}{ et al. }

\newcommand {\dz} {D^{0}}
\newcommand {\dzdzb} {\dz-\overline{\dz}}

\begin{document}
\footnotetext[0]{Received 29 May 2008, Revised 11 June 2008}
\title{A simulation study on the measurement of $\dzdzb$ mixing parameter y at BESIII
\thanks{Supported by National Natural Science Foundation of China (10491300,10491303,10735080),
Research and Development Project of Important Scientific Equipment
of CAS (H7292330S7), 100 Talents Programme of CAS (U-25, U-54,U-612)
and Scientific Research Fund of GUCAS(110200M202)} }


\author{
 HUANG Bin$^{1,2;1)}$\email{huangb@ihep.ac.cn}
\and ZHENG Yang-Heng$^{2;2)}$\email{zhengyh@gucas.ac.cn}
 \and LI Wei-Dong$^{1;3)}$\email{liwd@ihep.ac.cn}
 \and BIAN Jian-Ming$^{\ihep,\gscas}$
  \and CAO Guo-Fu$^{\ihep,\gscas}$
  \and CAO Xue-Xiang$^{\ihep,\gscas}$
  \and CHEN Shen-Jian$^{\nju}$
  \and DENG Zi-Yan$^{\ihep}$
  \and Fu Cheng-Dong$^{\tshu,\ihep}$
  \and GAO Yuan-Ning$^{\tshu}$
  \and HE Kang-Lin$^{\ihep}$
  \and HE Miao$^{\ihep,\gscas}$
  \and HUA Chun-Fei$^{\zzu}$
  \and HUANG Xing-Tao$^{\sdu}$
  \and JI Xiao-Bin$^{\ihep}$
  \and LI Hai-Bo$^{\ihep}$
  \and LIANG Yu-Tie$^{\pku}$
  \and LIU Chun-Xiu$^{\ihep}$
  \and LIU Huai-Min$^{\ihep}$
  \and LIU Qiu-Guang$^{\ihep}$
  \and LIU Suo$^{\lnu}$
  \and MA Qiu-Mei$^{\ihep}$
  \and MA Xiang$^{\ihep,\gscas}$
  \and MAO Ya-Jun$^{\pku}$
  \and MAO Ze-Pu$^{\ihep}$
  \and MO Xiao-Hu$^{\ihep}$
  \and PAN Ming-Hua$^{\gxnu}$
  \and PANG Cai-Ying$^{\gxnu}$
  \and PING Rong-Gang$^{\ihep}$
  \and QIN Gang$^{\ihep,\gscas}$
  \and QIN Ya-Hong$^{\zzu}$
  \and QIU Jin-Fa$^{\ihep}$
  \and SUN Sheng-Sen$^{\ihep}$
  \and SUN Yong-Zhao$^{\ihep,\gscas}$
  \and WANG Ji-Ke$^{\ihep,\gscas}$
  \and WANG Liang-Liang$^{\ihep,\gscas}$
  \and WEN Shuo-Pin$^{\ihep}$
  \and WU Ling-Hui$^{\ihep}$
  \and XIE Yu-Guang$^{\ihep,\gscas}$
  \and XU Min$^{\ustc}$
  \and YAN Liang$^{\ihep,\gscas}$
  \and YOU Zheng-Yun$^{\pku}$
  \and YU Guo-Wei$^{\ihep}$
  \and YUAN Chang-Zheng$^{\ihep}$
  \and YUAN Ye$^{\ihep}$
  \and ZHANG Chang-Chun$^{\ihep}$
  \and ZHANG Jian-Yong$^{\ihep}$
  \and ZHANG Xue-Yao$^{\sdu}$
  \and ZHANG Yao$^{\ihep}$
  \and ZHU Yong-Sheng$^{\ihep}$
  \and ZHU Zhi-Li$^{\gxnu}$
  \and ZOU Jia-Heng$^{\sdu}$
}
\maketitle
\address{ %
\ihep~(Institute of High Energy Physics, CAS, Beijing 100049, China）\\
 \gscas~（Graduate University of Chinese Academy of Sciences, Beijing 100049,China) \\
 \tshu~（Tsinghua University, Beijing 100084, China）\\
 \nju~ (Nanjing University, Nanjing 210093, China)\\
 \zzu~（Zhengzhou University, Zhengzhou 450001, China）\\
 \pku~（Peking University, Beijing 100871, China）\\
 \lnu~（Liaoning University, Shenyang 110036, China）\\
 \gxnu~（Guangxi Normal University, Guilin 541004, China）\\
 \ustc~（Department of Modern Physics, University of Science and Technology of China, Hefei 230026, China）\\
 \sdu~（Shandong University, Jinan 250100, China）\\
}

\begin{abstract}
We established a method on measuring the $\dzdzb$ mixing parameter
$y$ for BESIII experiment at the BEPCII $e^+e^-$ collider. In this
method, the doubly tagged $\psi(3770) \rightarrow D^0
\overline{D^0}$ events, with one $D$ decays to $CP$-eigenstates and
the other $D$ decays semileptonically, are used to reconstruct the
signals. Since this analysis requires good $e/\pi$ separation, a
likelihood approach, which combines the $dE/dx$, time of flight and
the electromagnetic shower detectors information, is used for
particle identification. We estimate the sensitivity of the
measurement of $y$ to be $0.007$ based on a $20fb^{-1}$ fully
simulated MC sample.
\end{abstract}



\footnotetext[0]{\hspace*{-2em}\small\centerline{\thepage\ --- \pageref{LastPage}}}%

\begin{multicols}{2}

\section{Introduction}

The mixing between a particle and its antiparticle has been observed
experimentally in neutral $K$, $B_d$ and $B_s$ system. In the
Standard Model, however, the mixing rate of the neutral $D$ system
is expected in general to be small and long-distance contributions
make it difficult to be
calculated\cite{ref:bigi,ref:burdman,ref:falk1,ref:falk2}. 
Recently, several
measurements\cite{ref:mixmeas1,ref:mixmeas2,ref:mixmeas3,ref:mixmeas4,ref:mixmeas5}
present evidences for $D^0-\bar{D^0}$ mixing with the significance
ranging from 3 to 4 standard deviations. This highlights the need
for independent measurements of the mixing parameters. Here, we
present a study on measuring the mixing parameter $y$ at BESIII
experiment, which takes advantage of the correlated threshold
production of $D^0-\bar{D^0}$ pairs in $e^+e^-$ collisions.

For the neutral $D$ meson system, two mass eigenstates and flavor
eigenstates are not equivalent and can be expressed as the following
form of the two quantum states:
\begin{equation} \displaystyle |D_{A,B} \rangle = p|D^{0} \rangle \pm
q|\bar{D^{0}}\rangle,
\end{equation}
with eigenvalues of masses and widths to be $m_{A,B}$ and
$\Gamma_{A,B}$. Conventionally, the $D^0-\bar{D^0}$ mixing is
described by two small dimensionless parameters:
\begin{equation}
\label{eqn:xypara} x \equiv \displaystyle\frac {\Delta m}{\Gamma}, y
\equiv \displaystyle\frac {\Delta \Gamma}{2\Gamma},
\end{equation}
where $\Delta m \equiv m_A - m_B$, $\Delta\Gamma \equiv \Gamma_A -
\Gamma_B$ and $\Gamma \equiv (\Gamma_A + \Gamma_B)/2$. The mixing
rate $R_{M}$ is approximately
\begin{equation}
R_{M} \approx \displaystyle\frac{x^{2}+y^{2}}{2}.
\end{equation}
In the limit of $CP$ conservation, the $|D_{A}\rangle$ and
$|D_{B}\rangle$ denote the $CP$ eigenstates.

The mixing parameters can be measured in several ways. The
$B$-factories measured $R_M$ with semileptonic $D^0$ decay
samples\cite{ref:mixmeas6,ref:mixmeas7}.
Reference\cite{ref:mixmeas8,ref:mixmeas9} also gave an estimation on
the sensitivities of $R_M$ measurement at BESIII. Other
attempts\cite{ref:mixmeas1,ref:mixmeas2,ref:mixmeas3,ref:mixmeas4,ref:mixmeas5}
are based on the proper-time measurements of the neutral $D$ meason
decays. However, the time-dependent analyse are not possible at
symmetric charm factory, which operates at the $\psi(3770)$
resonance. In this analysis, we utilize the quantum-coherent
threshold production of $\dzdzb$ pairs in a state of definite $C =
-1$. Applying the kinematics of the process of $e^+e^- \rightarrow
\psi(3770) \rightarrow D^0 \bar{D^0}$, we can reconstruct both
neutral $D$ mesons (double tagging (DT) technique) to obtain clean
samples to measure the mixing parameters, the strong phase
difference and the $CP$ violation. For the single $D^0$ meson decays
into a $CP$ eigenstate, the time-integrated decay rate can be
written as\cite{ref:asner1,ref:asner2}:
\begin{equation}
\Gamma_{CP\pm} \equiv \Gamma \left( D^0 \rightarrow f_{CP\pm}\right)
= 2 A^2_{CP\pm}\left[1 \mp y \right],
\end{equation}
where $f_{CP\pm}$ is a $CP$ eigenstate with eigenvalue $\pm 1$, and
$A_{CP\pm} \equiv |\langle f_{CP\pm}| {\cal H}|D^0 \rangle |$ is the
magnitude of decay amplitude. If we consider the coherent $D$-pair
decays, in which one $D$ decays into $CP$ eigenstates and the other
$D$ decays semileptonically, the decay rate of $\left( D^0
\overline{D^0}\right)^{C=-1} \rightarrow \left( l^{\pm} X \right)
\left( f_{CP\pm} \right)$ is described
as\cite{ref:gronau,ref:xing1,ref:xing2}:
\begin{equation}
\Gamma_{l;CP} \equiv \Gamma \left[\left(
l^{\pm}X\right)\left(f_{CP}\right)\right] \approx A^2_{l^{\pm}X}
A^2_{CP},
\end{equation}
where $A_{l^{\pm}X} \equiv |\langle l^{\pm}X | {\cal H}|D^0 \rangle
|$. Here, we neglect terms to order $y^2$ or higher since $y$ is
much smaller than unit. We, thus, can derive:
\begin{equation}
y = {1 \over 4} \left(
\frac{\Gamma_{l;CP+}\Gamma_{CP-}}{\Gamma_{l;CP-}\Gamma_{CP+}} -
\frac{\Gamma_{l;CP-}\Gamma_{CP+}}{\Gamma_{l;CP+}\Gamma_{CP-}}
\right).
\end{equation}

To measure $y$ at BESIII, only the electron channels are used to
reconstruct the semileptonic $D^0$ decays. In the muon channels, the
transverse momentum of muon is too low to be efficiently identified
by the BESIII muon detector. Thus, the $e/\pi$ separation plays an
essential role to suppress the backgrounds. Fig.~\ref{fig:e_mom}
shows the momentum distribution of the electrons from the
semileptonic $D$ decays. The momentum distribution of the pions from
$s$ quark decays is similar to Fig.~\ref{fig:e_mom}. As a result,
the performance of electron identification (e-ID) will determine the
precision of the measurement of $y$ parameter.

\begin{center}
\includegraphics[width=8cm,height=4.8cm]{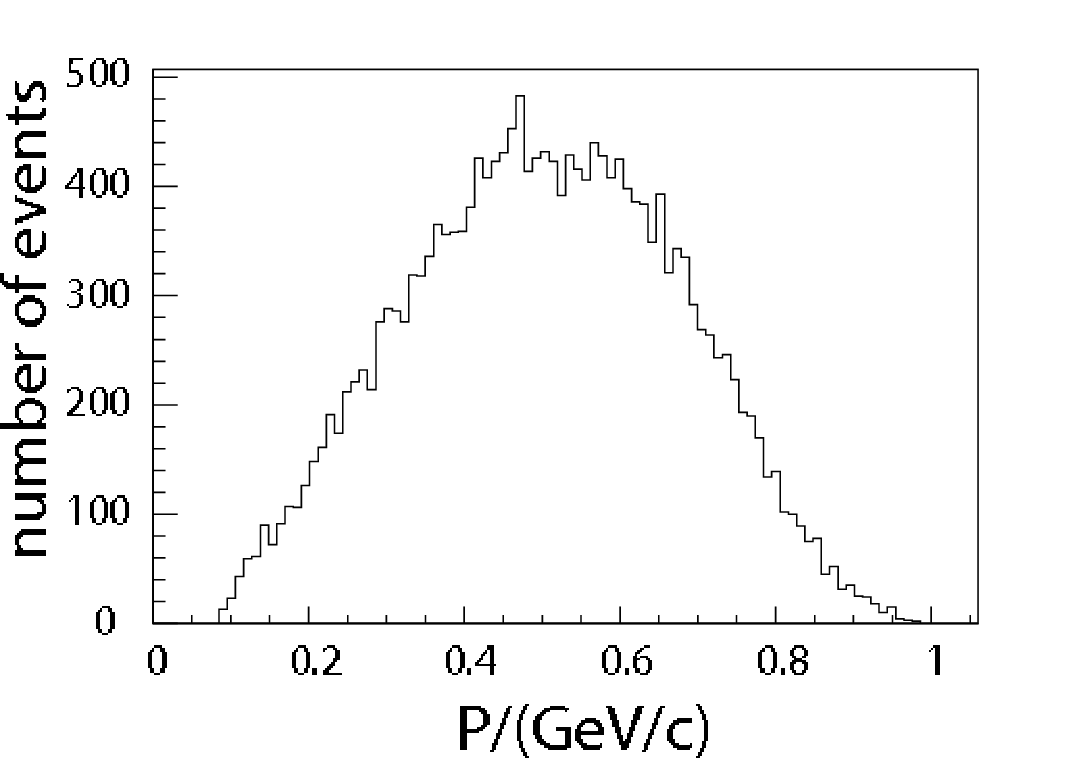}
\figcaption{Momentum of the electron from $D^{0}$ semileptonic
decays} \label{fig:e_mom} \end{center}

The designed peak luminosity of BEPCII (Beijing Electron Position
Collider) is $10^{33}cm^{-2}s^{-1}$ at beam energy $E_{beam}$ = 1.89
GeV, which is the highest in the tau-charm region ever planned and
an unprecedented large number of $\psi(3770)$ events is expected.

This paper is organized as follows: an improved electron
identification technique for BESIII is described in Section 2. In
Section 3, we describe the method on reconstructing the signals with
Mente Carlo(MC) simulation samples. Section 4 presents the estimated
sensitivity of $y$ measurement. The summary is presented in Section
5.

\section{Electron identification} \label{sec:pid}
The BESIII detector operates at BEPCII and consists of a beryllium
beam pipe, a helium-based small-celled drift chamber, Time-Of-Flight
(TOF) counters for particle identification, a CsI(Tl) crystal
electromagnetic calorimeter (EMC),  a super-conducting solenoidal
magnet with the field of 1 Tesla, and a muon identifier of Resistive
Plate Counters (RPC) interleaved with the magnet yoke plates. The
BESIII Offiline Software System (BOSS)\cite{ref:boss} of version
6.1.0 is used for this analysis. The detector
simulation\cite{ref:boost} is based on GEANT4\cite{ref:geant4}.

The BESIII detector has four subsystems for particle identification:
the dE/dx of the main drift chamber(MDC), TOF, EMC and the muon
counter. Among them, the dE/dx and the TOF systems are mainly used
for hadron separation, the EMC provides information for electron and
photon identification, the MUC has good performance on muon
identification\cite{ref:bes3_det_tdr}.

For electron identification, Refs\cite{ref:pid,ref:bes3_tof}
illustrate the use of dE/dx of MDC and TOF information. Here, an
improved $e/\pi$ separation technique is introduced in the following
sections.

\subsection{Electromagnetic calorimeter}

The BESIII electromagnetic
calorimeter\cite{ref:bes3_det_tdr,ref:bes3_det} is composed of one
barrel and two endcap sections, covering 93\% of 4$\pi$. There are a
total of 44 rings of crystals along the z direction in the barrel,
each with 120 crystals. And there are 6 layers in the endcap, with
different number of crystals in each layer. The entire calorimeter
has 6240 CsI(Tl) crystals with a total weight of about 24 tons. The
energy resolution is expected to be 2.5\% and the spatial resolution
is expected to be 0.6 cm for 1 GeV/c photon.

The primary function of the EMC is to precisely measure the energies
and positions of electron and photon. In order to distinguish
electron from hadron, we make use of significant differences in
energy deposition and the shower shape of different type of the
particles.

\subsection{Variables used in e-ID}

The following variables are used to identify the electron from pion:
\begin{description}
\item[1)] Ratio of the energy measured by the EMC and the momentum of the charged track by the
MDC ($E/p$).
\end{description}
Ratio of the energy measured by the EMC and the momentum of the
charged track by the MDC ($E/p$). When an electron passes through
the calorimeter, the electron produces electromagnetic shower and
loses its energy by pair-production, Bremsstrahlung and
ionizing/exciting atomic electrons. Since the mass of electron is
negligible in the energy range of interest, we expect to have the
ratio $E/p=1$ within the measurement errors. For hadrons, the $E/p$
is typically smaller than one.
\begin{description}
  \item[2)] Lateral shower shape at the EMC.
\end{description}
In order to enhance the separation between the electrons and the
interacting hadrons, the lateral shower shape can also be utilized.
These variables include: $E_{seed}/E_{3\times3},
E_{3\times3}/E_{5\times5}$ and the second-moment. Here the
$E_{seed}$ is the energy deposited in the central crystal, the
$E_{3\times3}$ and $E_{5\times5}$ represent the energy deposit in
the $3\times3$ and $5\times5$ crystal array, respectively. The
second-moment $S$ is defined as
\begin{equation}
\displaystyle S=\frac{\sum_{i}{E_{i}\cdot
d_{i}^{2}}}{\sum_{i}{E_{i}}},
\end{equation}
where $E_{i}$ is the energy deposit in the  $i$-th crystal, and
$d_{i}$ is the distance between the $i$-th crystal and the center
position of reconstructed shower. Detailed description of $E/p$ and
the lateral shower shape can be found in Ref~\cite{ref:pid}.
\begin{description}
  \item[3)] Longitudinal shower shape at the EMC.
\end{description}
The longitudinal shower shape provides additional information for
electron identification. The variable $\Delta\phi$, between the
polar angles where the track intersects the EMC and the shower
center, can be used. The distributions of $\Delta\phi$
 for electron and pion are drawn in Fig.~\ref{fig:dleta_phi}. The center of
electron showers is closer to the impact point of track on EMC since
the electron showers reach their maximum earlier than hadrons.
\begin{center}
\includegraphics[width=3.8cm,height=2.8cm]{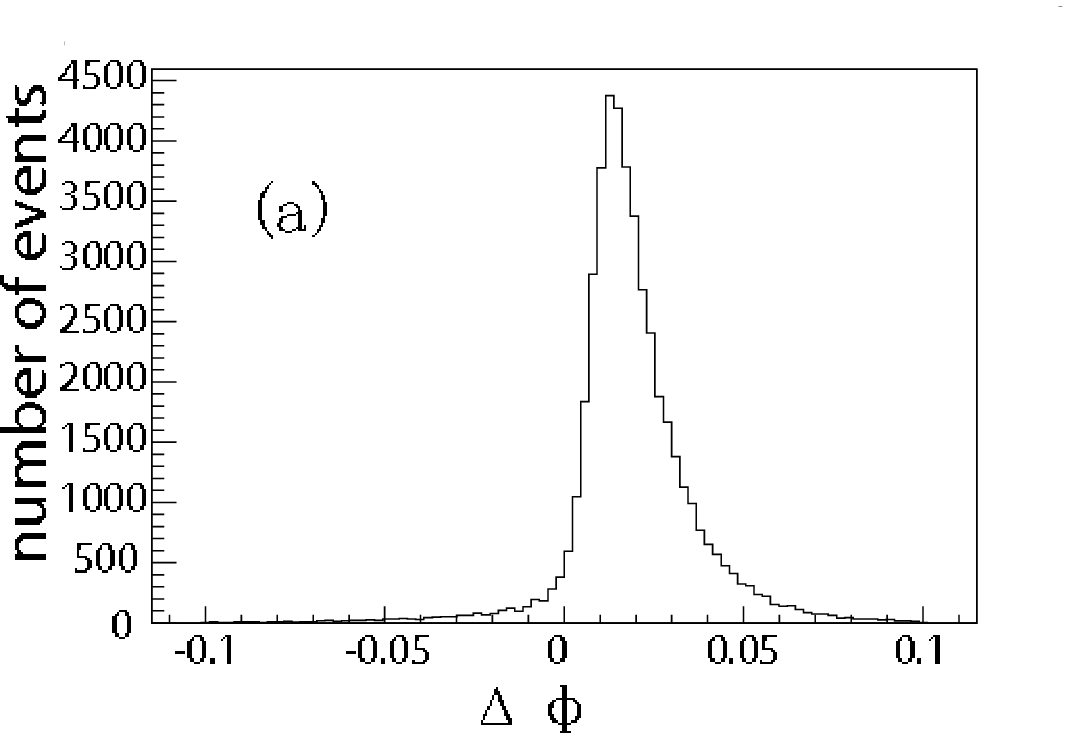}\quad
\includegraphics[width=3.8cm,height=2.8cm]{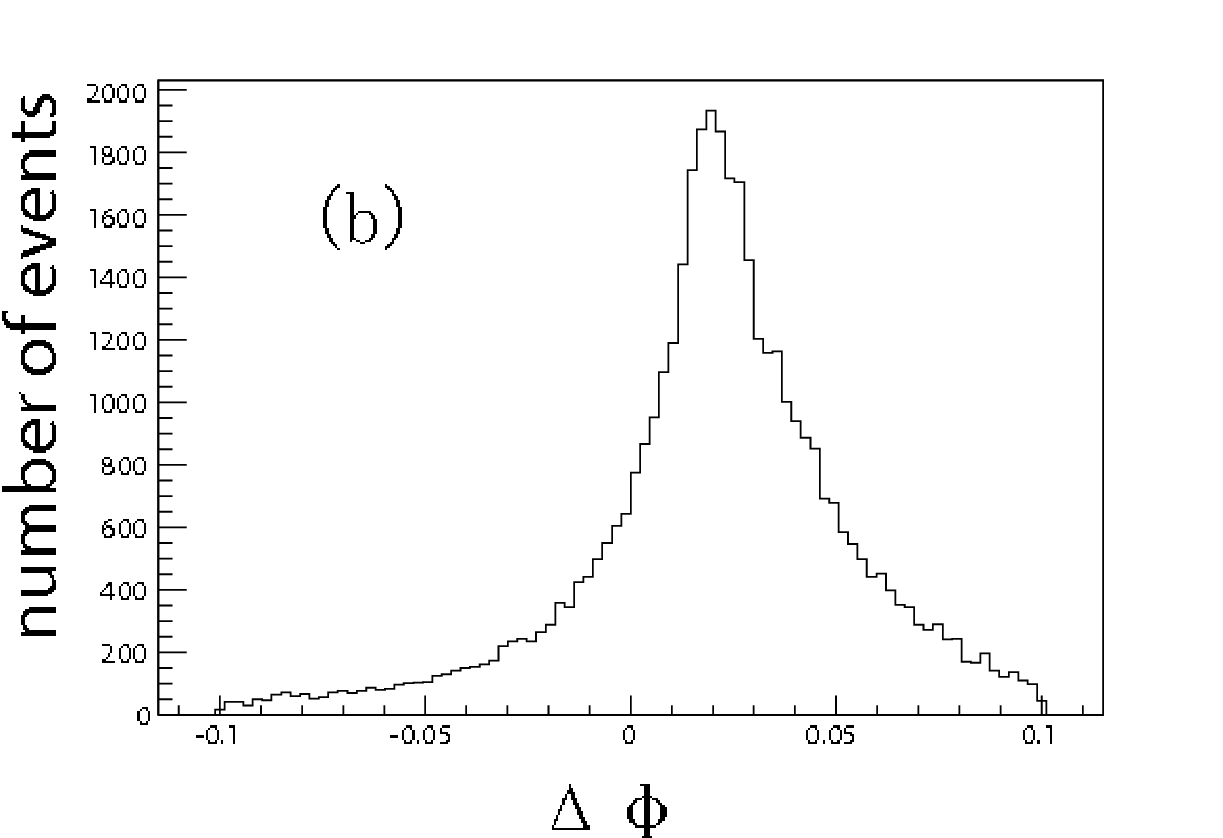}\\
\figcaption{$\Delta\phi$ of (a) electron (b) pion.}
\label{fig:dleta_phi}
\end{center}

\subsection{The correlation between variables}

The $E/p$ ratio, lateral shower shape and longitudinal shower shape
are all depending on the deposited energy in the crystals. Thus,
these variables may be correlated. We calculate the correlation
coefficients $\rho_{ij}$ between the $E/p$,
$E_{3\times3}/E_{5\times5}$ and $\Delta\phi$ using the function:
\begin{equation} \displaystyle
M_{i,j}=\sum_{i,j}{(x_{i}-\bar{x_i})\cdot(x_{j}-\bar{x_j})},\quad
\rho_{ij}=\frac{M_{ij}}{\sqrt{M_{ii}\times
M_{jj}}},\label{eq:Correlation parameter}
\end{equation}
where $i,j$ are the indices of the variable names.
Figure.~\ref{fig:cor} shows the correlation between any two of the
variables of electron and pion, respectively, with the momentum
ranging from 0.2 GeV/c to 2.0 GeV/c. Here, the x-axis represents the
particle momentum and the y-axis represents the correlation
coefficient $\rho_{ij}$. The distribution indicates strong
correlation between the variables.
\begin{center}
\includegraphics[width=2.8cm,height=2.1cm]{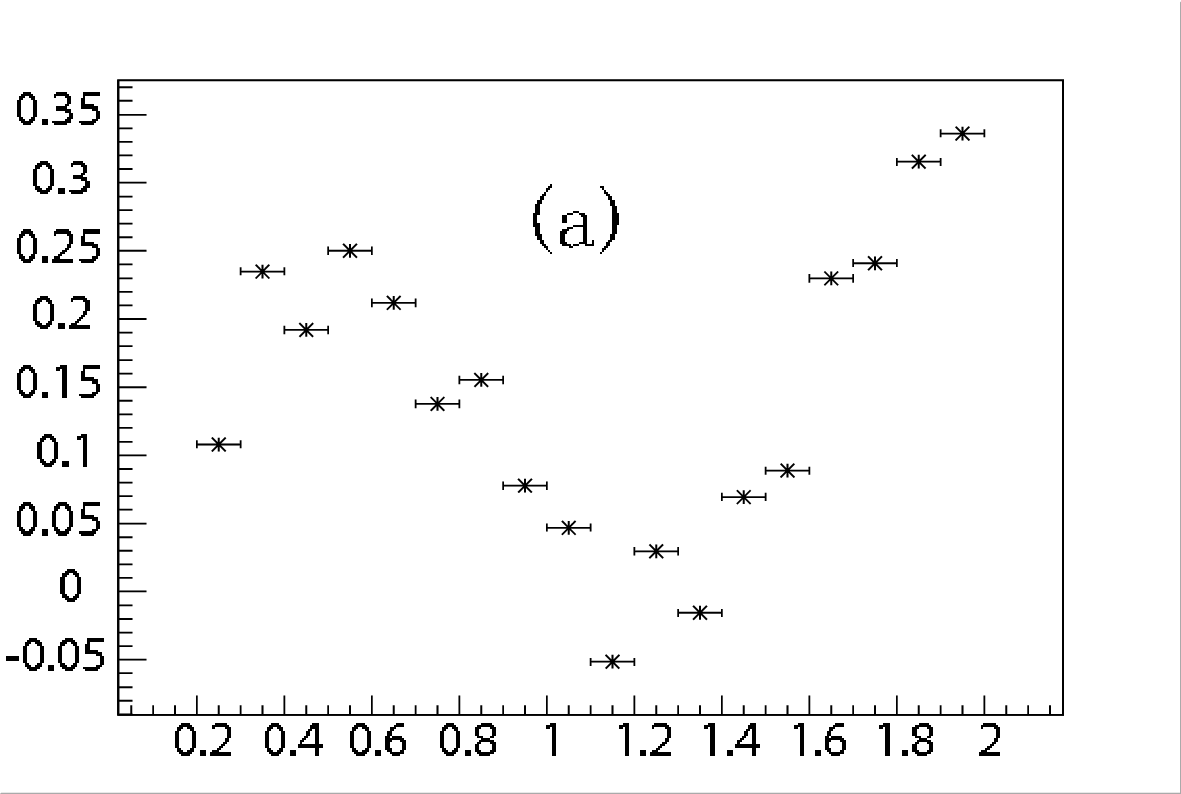}\includegraphics[width=2.8cm,height=2.1cm]{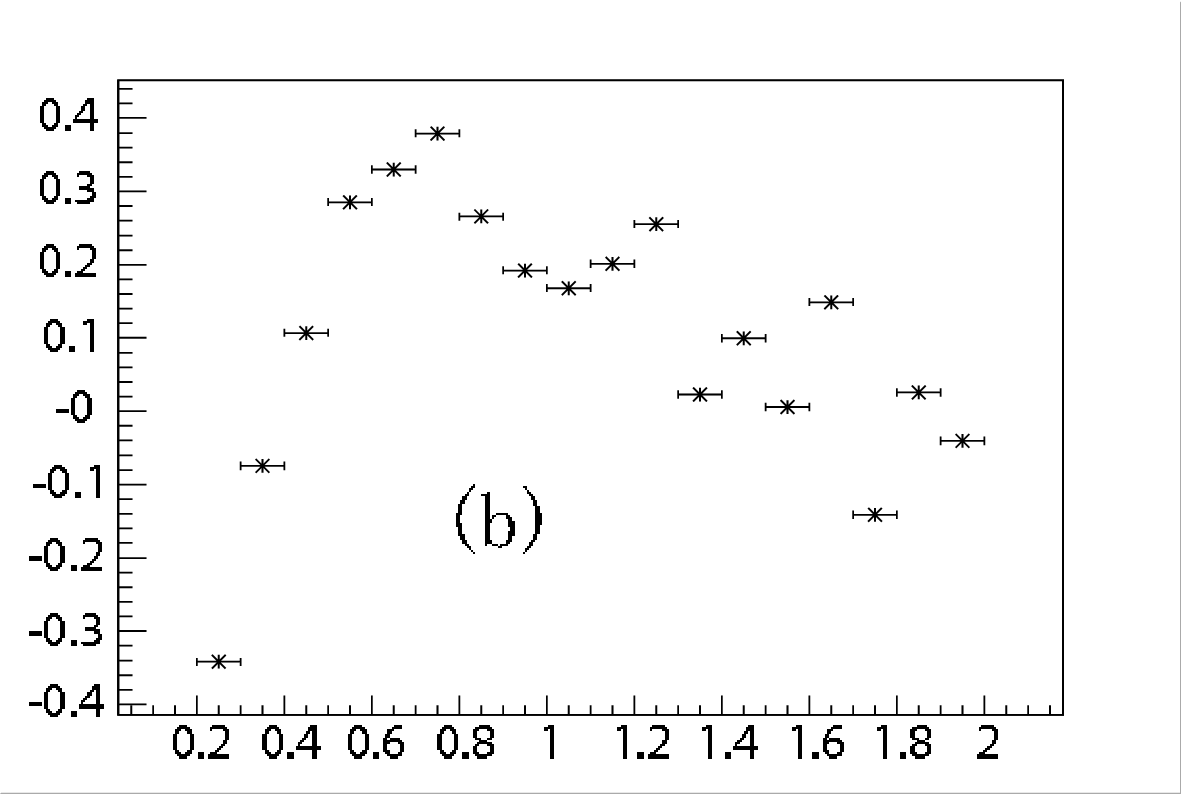}\includegraphics[width=2.8cm,height=2.1cm]{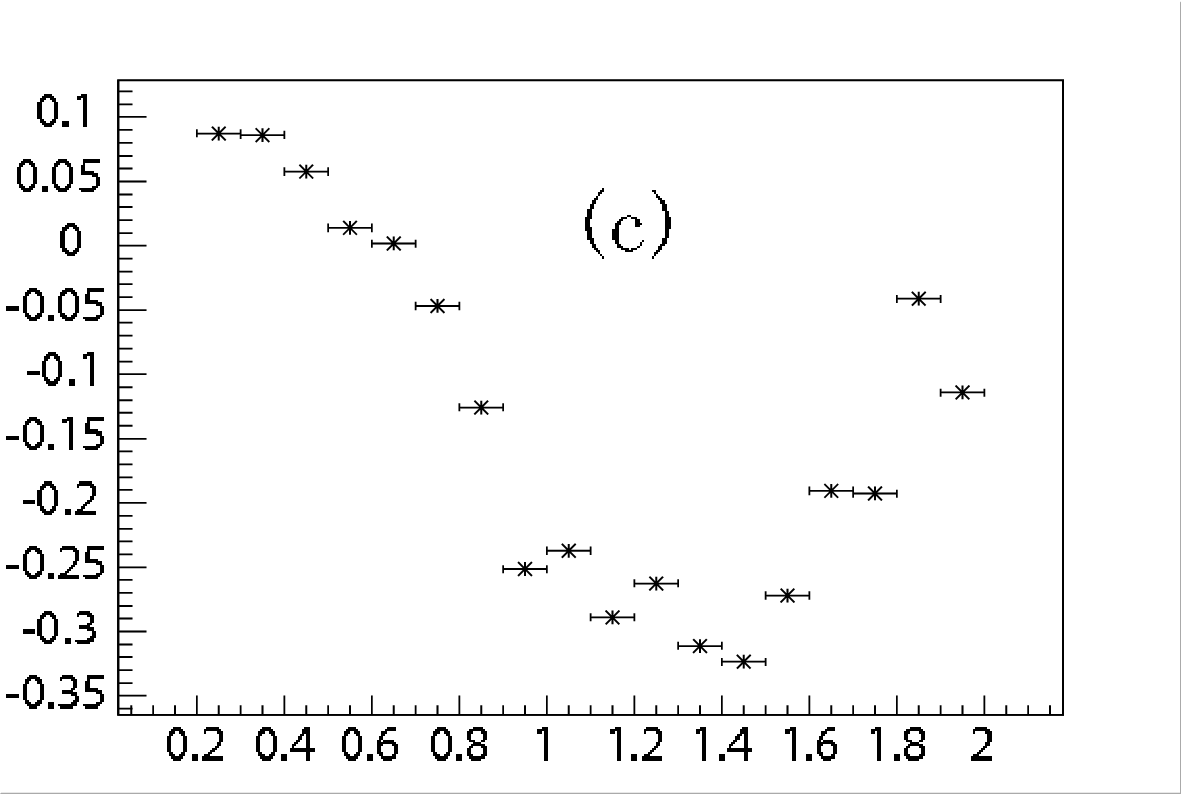}\quad
\includegraphics[width=2.8cm,height=2.1cm]{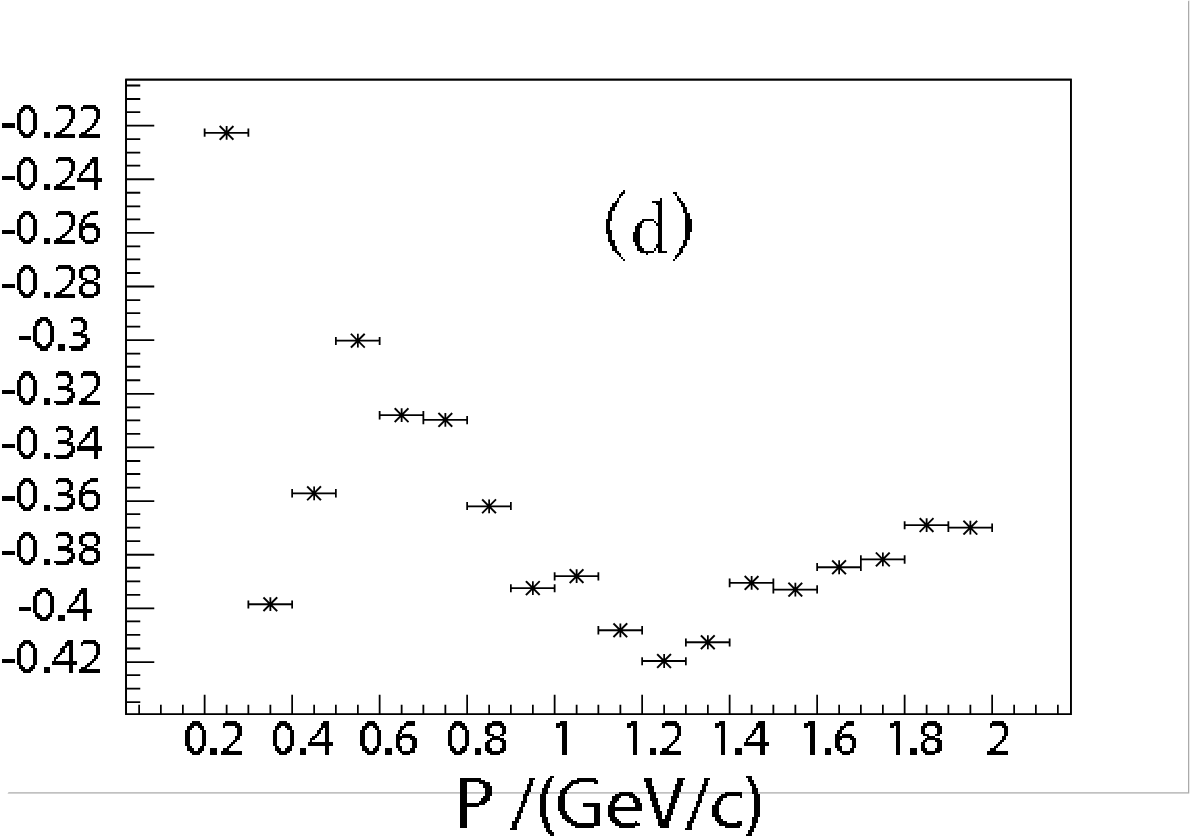}\includegraphics[width=2.8cm,height=2.1cm]{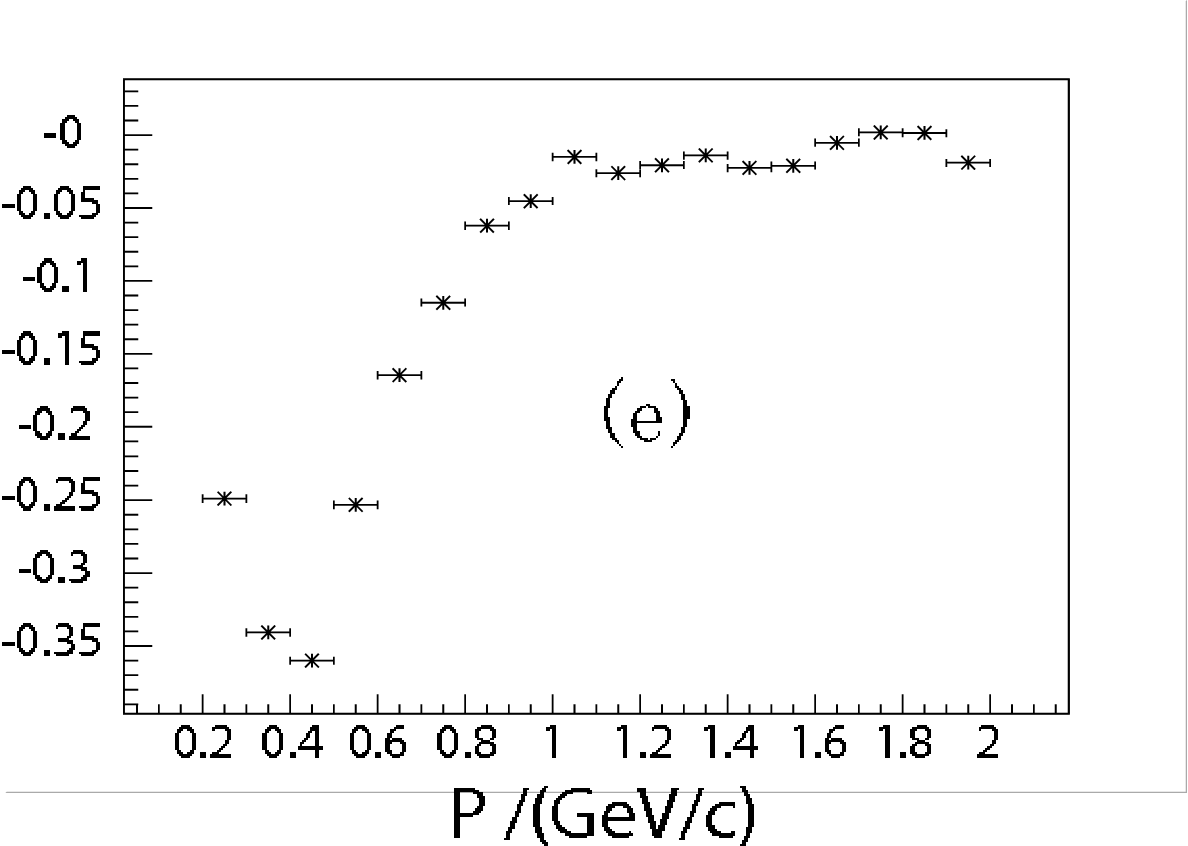}\includegraphics[width=2.8cm,height=2.1cm]{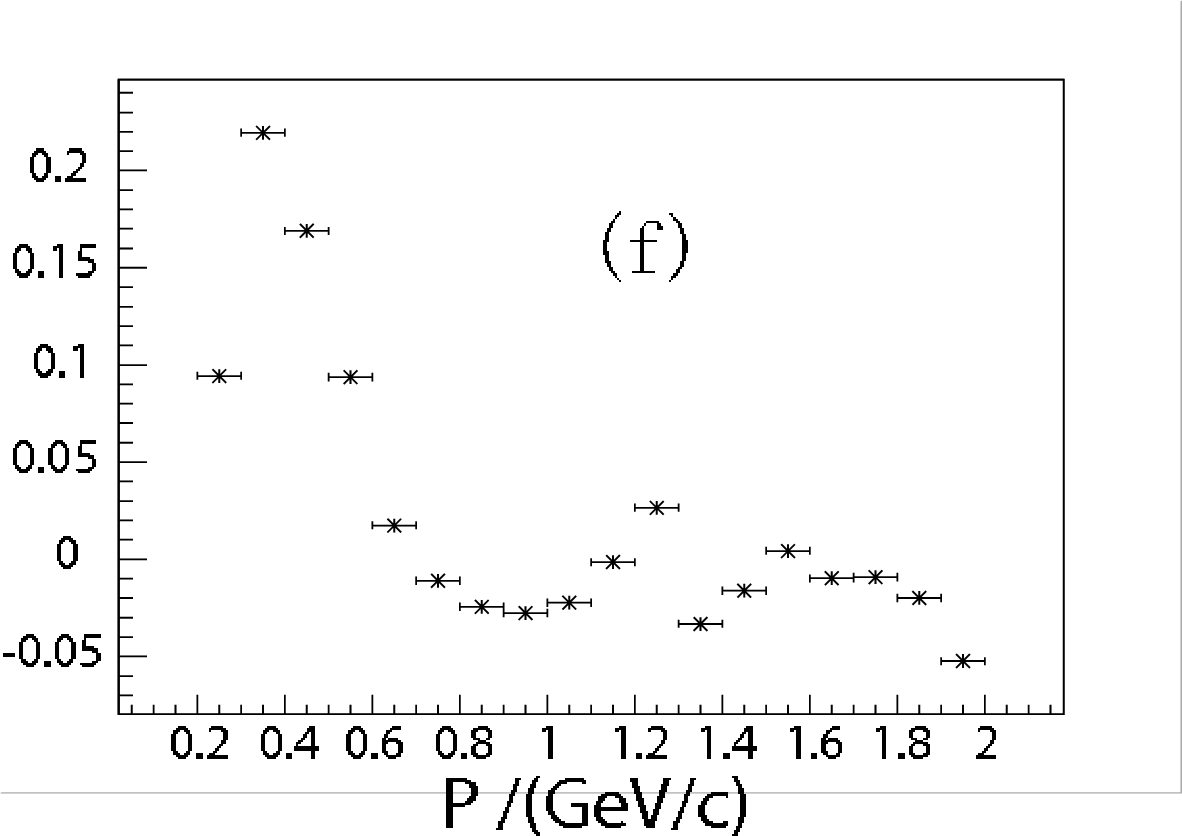}\quad
\figcaption{Correlations between (a)E/p and $E_{seed}/E_{3\times3}$
of electron; (b)E/p and $E_{3\times3}/E_{5\times5}$ of electron;
(c)$E_{seed}/E_{3\times3}$ and $E_{3\times3}/E_{5\times5}$ of
electron; (d)E/p and $E_{seed/E3\times3}$ of pion; (e)E/p and
$E_{3\times3}/E_{5\times5}$ of pion; (f)$E_{seed}/E_{3\times3}$ and
$E_{3\times3}/E_{5\times5}$ of pion.} \label{fig:cor}
\end{center}

\subsection{PID Algorithm}
Considering the correlations between the variables, the traditional
method for particle identification may be underperforming. In the
e-ID, we implement the artificial neural network
(ANN)~\cite{ref:pid_ann} to provide a general framework for
estimating non-linear functional mapping between the input variables
and the output variable. For the neural network (NN) training, we
use the momentum, traverse momentum and other six discriminants
(total deposit energy, Eseed, E3x3, E5x5, second moment and
$\Delta\phi$) as the input variables. The network we choose has one
hidden layer with 16 neurons and one output value.
Figure.~\ref{fig:NN outputs} shows the two-dimension distributions
of the output value versus the momentum of the electrons and pions.
It is obvious that the distribution of the output value depends on
the momentum, especially at low momentum region. Thus, it is
unsuitable to apply a single cut on the output value to separate the
electrons from the pions. In practice, we construct probability
density function (PDF) of the NN output value at every 0.1 GeV/c
momentum bin. The PDF is obtained from fitting the nearest 4 bins of
the NN output value, with the third-order polynomial function. Then,
the PDF value of the NN output can be extracted from the fit.
Finally, we make the PID decision by comparing the likelihood values
of electron and pion hypothesis.
\begin{center}
\includegraphics[width=8cm,height=6cm]{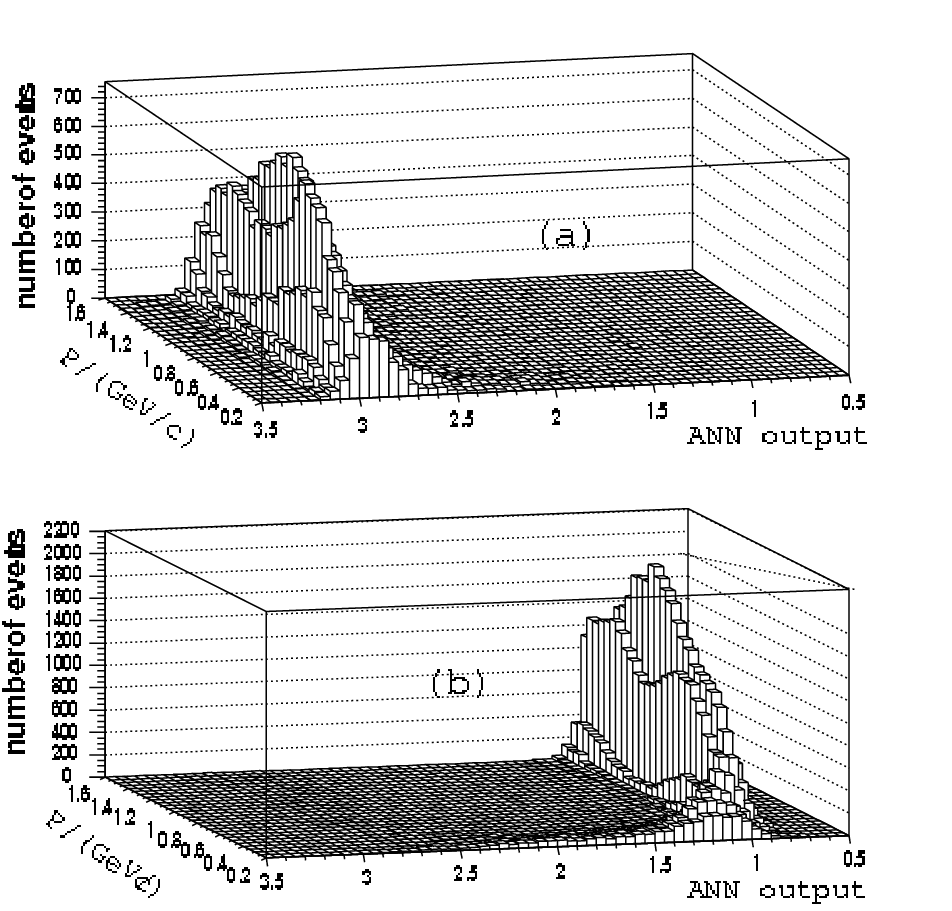}\\
\figcaption{The NN outputs of (a) pion (b) electron
 samples with the momentum ranging from 0.2GeV/c to 1.6GeV/c} \label{fig:NN outputs}
\end{center}

\subsection{The performance check}
To combine the $dE/dx$, TOF and EMC information, the likelihood
approach\cite{ref:pid_likelihood} is adopted. Firstly, the
likelihood value of each subsystem is calculated. Then, the total
likelihood value of each hypothesis is calculated by the following
formula:
\begin{equation} \displaystyle
L_{tot}=L_{dE/dx}*L_{TOF}*L_{EMC},
\end{equation}
where $L_{dE/dx}$ and $L_{TOF}$ represent the likelihood value of
$dE/dx$ and TOF subsystems respectively. Finally, the likelihood
ratio of electron hypothesis is defined as:
\begin{equation} \displaystyle
lhf_{_e}=\frac{L_{e}}{L_{e}+L_{\pi}},
\end{equation}
where $L_{e}$ and $L_{\pi}$ are the total likelihood value of
electron and $\pi$ hypothesis. To check the performance of the
$e/\pi$ separation, both the electron and pion samples are generated
with the momentum ranging from 0.2 Gev/c to 1.6 GeV/c, by using
single particle generator. Fig.~\ref{fig:dis}(a) shows the electron
likelihood ratio distributions of these samples. For a particle to
be identified as an electron, we require $lhf_{_e}>0.5$.
Fig.~\ref{fig:dis}(b) shows the combined $e/\pi$ separation
performance using the $dE/dx$, TOF and EMC systems.

\begin{center}
\includegraphics[width=8cm,height=4.8cm]{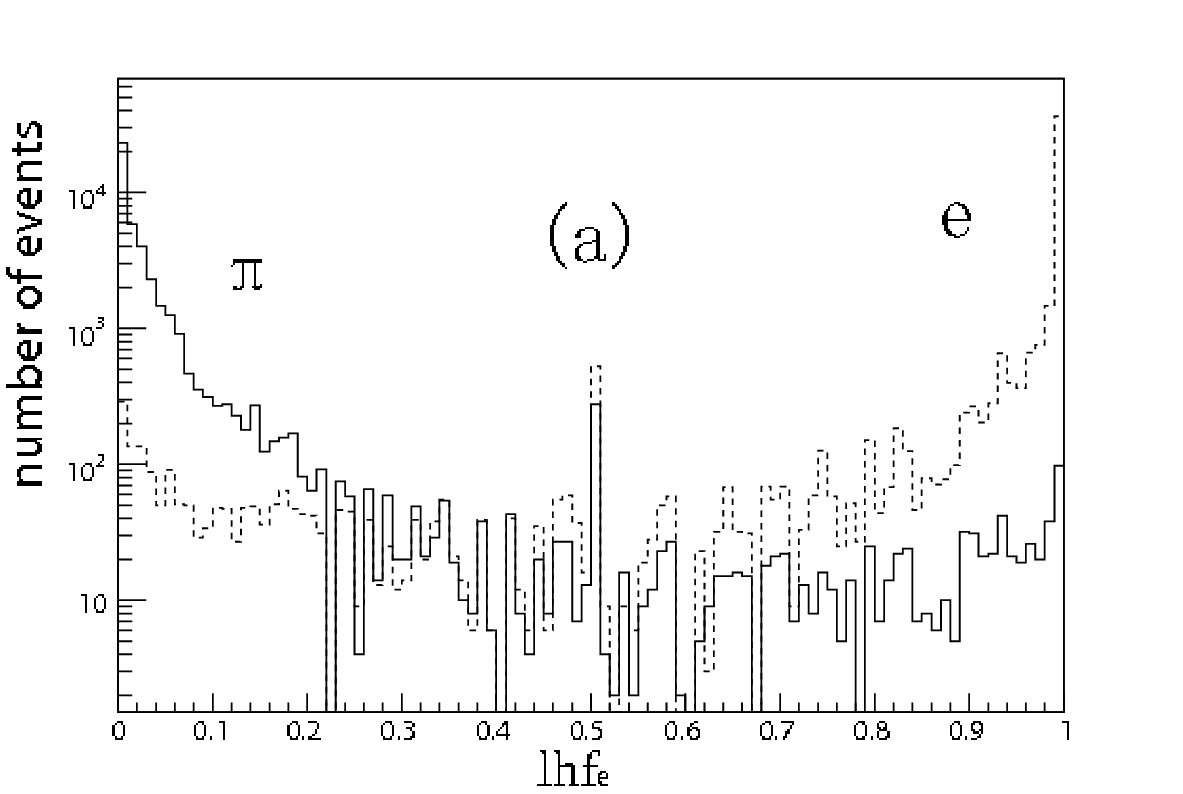}\\
\includegraphics[width=8cm,height=4.8cm]{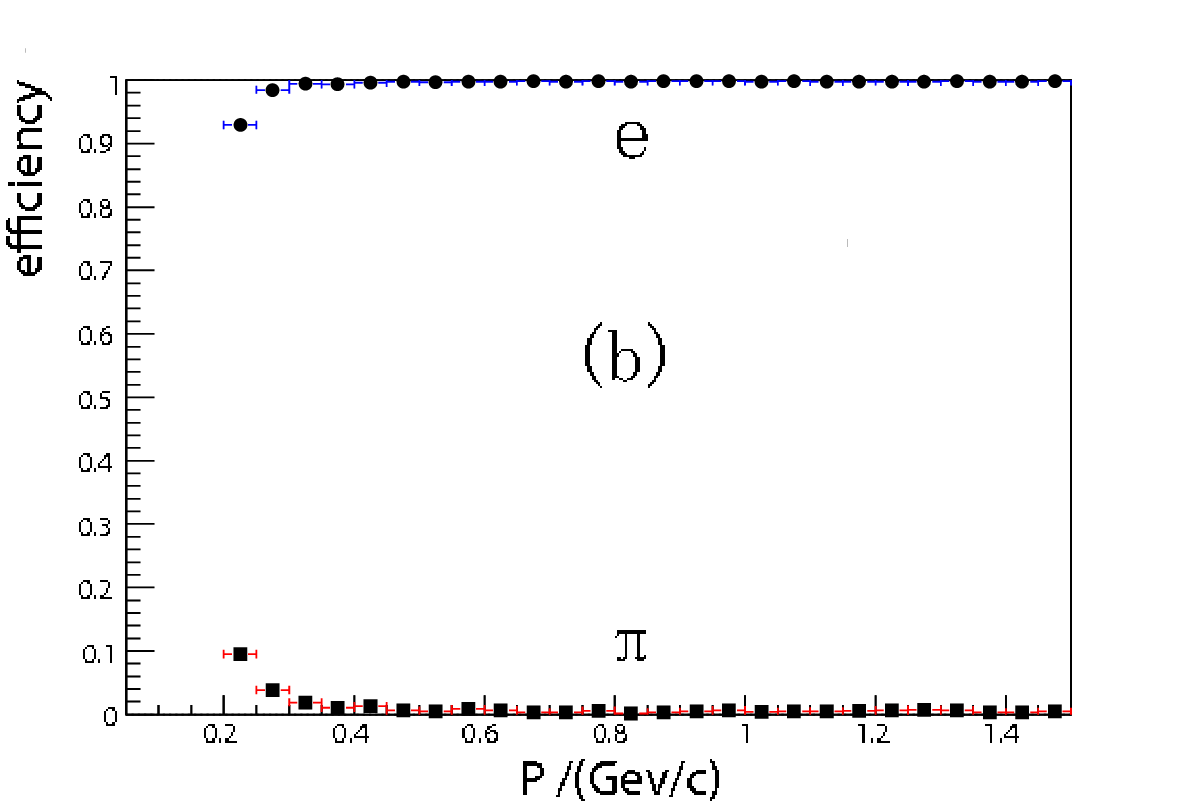}\\
\figcaption{ (a)$lhf_{_e}$ of electron and $\pi$ samples; (b)
performance of $e/\pi$ seperation.} \label{fig:dis}
\end{center}

\section{Simulation and reconstruction} \label{sec:rec}
\subsection{The reconstruction of $CP$ tags}
\label{sec:CP tag}

For the neutral $D$ meson decays, the main decay modes of $CP+$
eigenstate are $K^{+}K^{-}$, $\pi^{+}\pi^{-}$,
$K_{S}\pi^{0}\pi^{0}$, $\pi^{0}\pi^{0}$, $K_{S}K_{S}$ and
$\rho^{0}\pi^{0}$. The $CP-$ eigenstates decay through the modes
$K_{S}(\pi^{0}, \rho^{0}, \eta, \eta', \phi, \omega)$. Considering
the branching ratio and the reconstruction efficiency, we only
simulated the $K^{+}K^{-}$, $\pi^{+}\pi^{-}$ for $CP+$ tagging, and
the $K_{S}(\pi^{0}, \eta, \eta')$ for $CP-$ tagging.

For selecting the charged tracks, the following selection criteria
are adopted:
\begin{description}
  \item[1)] All charged
tracks must have a good helix fit, and are required to be measured
in the fiducial region of MDC;
  \item[2)]  Their parameters must be
corrected for energy loss and multiple scattering according to the
assigned mass hypotheses;
  \item[3)] The tracks not associated with
$K^{0}_{S}$ reconstruction are required to be originated from the
interaction point(IP).
\end{description}

For reconstructing the $CP+$ eigenstates, two opposite-charged
tracks of $K$ or $\pi$ are selected with the requirements that they
are from IP and to pass a common vertex constraint. To identify a
track as a $\pi$ or $K$, we use the likelihood method to combine the
information of $dE/dx$ and TOF with the likelihood fraction of $\pi$
or $K$ greater than 0.5. Then the beam constrained mass($M_{bc}$) of
the $D$ meson is used to distinguish the signal and background, and
it is defined as:
\begin{equation} \displaystyle
M_{bc}\equiv\sqrt{E_{beam}^{2}-(\sum\bm{p}_{i})^{2}}=\sqrt{E_{beam}^{2}-(\bm{p}_{D})^{2}},
\end{equation}
where the $E_{beam}$ is the beam energy, the $\bm{p}_{i}$ is the
momentum of the $i$-th track and the $\bm{p}_{D}= \sum\bm{p}_{i}$ is
the momentum of the reconstructed $D$ meson.

For tagging the $CP-$ eigenstates, we need to reconstruct the
neutral mesons $K_{S}$, $\pi^{0}$, $\eta$ and $\eta^{'}$. The
$K_{S}$ candidates are reconstructed through the decay of $K_{S}
\rightarrow\pi^{+}\pi^{-}$. The decay vertex formed by
$\pi^{+}\pi^{-}$ pair is required to be away from the interaction
point, and the momentum vector of $\pi^{+}\pi^{-}$ pair must be
aligned with the position vector of the decay vertex to the IP. Here
we set $L_{vtx}/\sigma_{vtx}$ to be greater than 2, where $L_{vtx}$
and $\sigma_{vtx}$ are the measured decay length and the error of
the decay length of the $K_S$. The $\pi^{+}\pi^{-}$ invariant mass
is required to be consistent with the $K_{S}$ nominal mass within
$\pm10$ MeV. To identify the neutral tracks, one has to address a
number of processes which can produce both real and spurious showers
in EMC. The major source of these ``fake photons" arises from
hadronic interaction, which can create a ``split-off" shower. This
shower does not associate with the main shower and may be recognized
as a photon. Other sources of fake photons include particle decays,
back splash, beam associated background and electronic noise. To
reject ``fake photons", the selection criteria for ``good photon"
include a deposit energy cut, and a spatial cut, which requires that
the cluster is isolated from the nearest charged tracks. These
``cuts" are set to be $E_{\gamma}>40MeV$ and
$\Delta_{c\gamma}>18^{\circ}$, where $E_{\gamma}$ and
$\Delta_{c\gamma}$ represent the deposited energy and the crossing
angle of the cluster to the nearest charged track, respectively. The
neutral pions are reconstructed from
$\pi^{0}\rightarrow\gamma\gamma$ decays using the photons observed
in the barrel and endcap regions of EMC. At the energies of
interest, a $\pi^{0}$ decays into two isolated photons. In addition,
we also reconstruct $\eta/\eta^{'}$ candidates in the modes of
$\eta\rightarrow\gamma\gamma,
\eta\rightarrow\pi^{+}\pi^{-}\pi^{0},\eta^{'}\rightarrow\gamma\rho^{0}
$ and $\eta^{'}\rightarrow\eta\pi^{+}\pi^{-}$. For these modes,
3$\sigma$ consistency with the $\pi^{0}/\eta/\eta^{'}$ mass is
required, followed by a kinematic mass constraint. For $CP-$
eigenstates, the beam constrained mass is also used to select the
signal.

Under the environment of BOSS 6.1.0, we simulated $\dzdzb$ pairs
production at the $\psi(3770)$ peak with one $D$ decayed into $CP$
eigenstates and the other $D$ decayed semileptonically. The $CP+$
eigenstates are decayed through $\pi^{+}\pi^{-}$ and $K^{+}K^{-}$
according to their branching ratios. For $CP-$ eigenstates, the
decay modes $K_{S}\pi^{0}$, $K_{S}\eta$ and $K_{S}\eta^{'}$ are
included. In the $K_{S}$, $\pi^{0}$, $\eta$ and $\eta^{'}$ decays,
the decay modes are listed as follows:
$K_{S}\rightarrow\pi^{+}\pi^{-}$, $\pi^{0}\rightarrow\gamma\gamma$,
$\eta\rightarrow\gamma\gamma$,
$\eta\rightarrow\pi^{+}\pi^{-}\pi^{0}$,
$\eta^{'}\rightarrow\gamma\rho^{0}$, and
$\eta^{'}\rightarrow\eta\pi^{+}\pi^{-}$. For the $CP+$ and $CP-$
eigenstates, we generated 30,000 events for each MC sample. The
distributions of the beam constrained mass of $D$ meson are shown in
Fig.~\ref{fig:mass CPm}.

\begin{center}
\includegraphics[width=3.8cm,height=2.8cm]{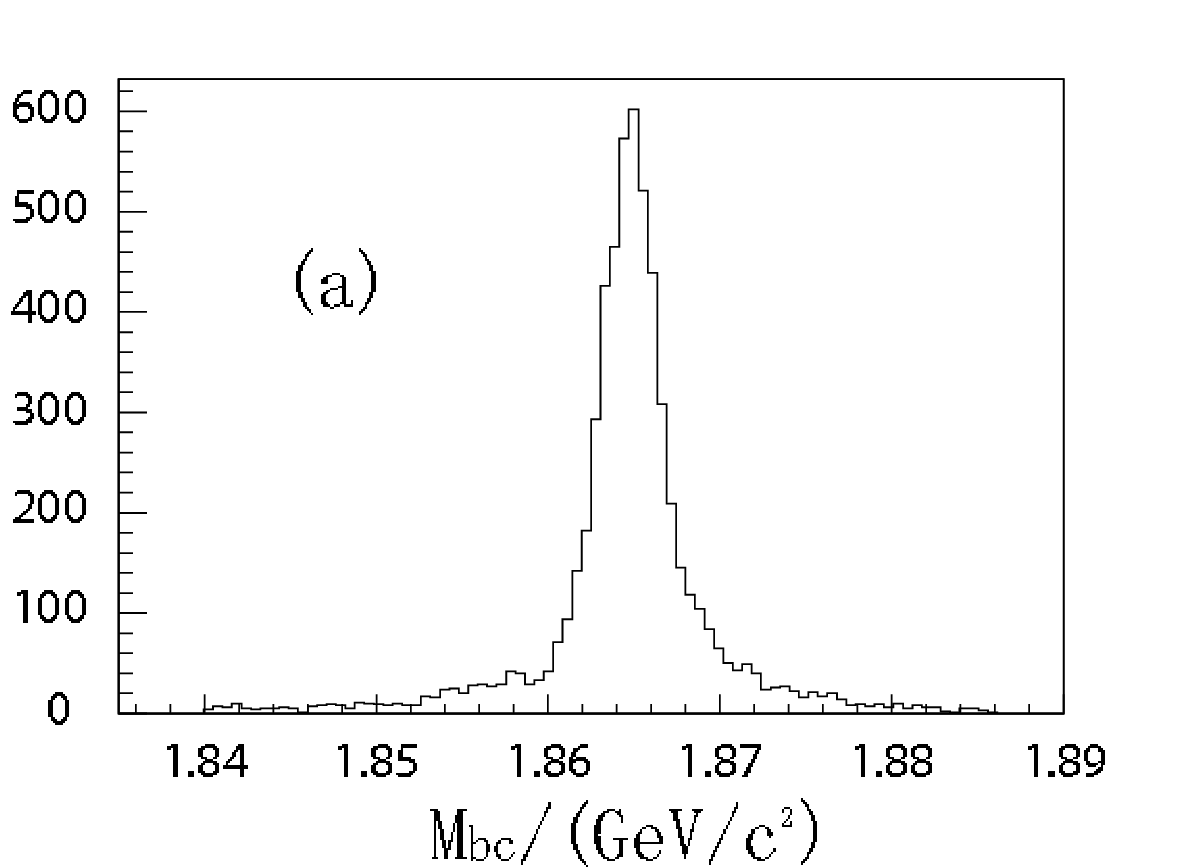}\quad
\includegraphics[width=3.8cm,height=2.8cm]{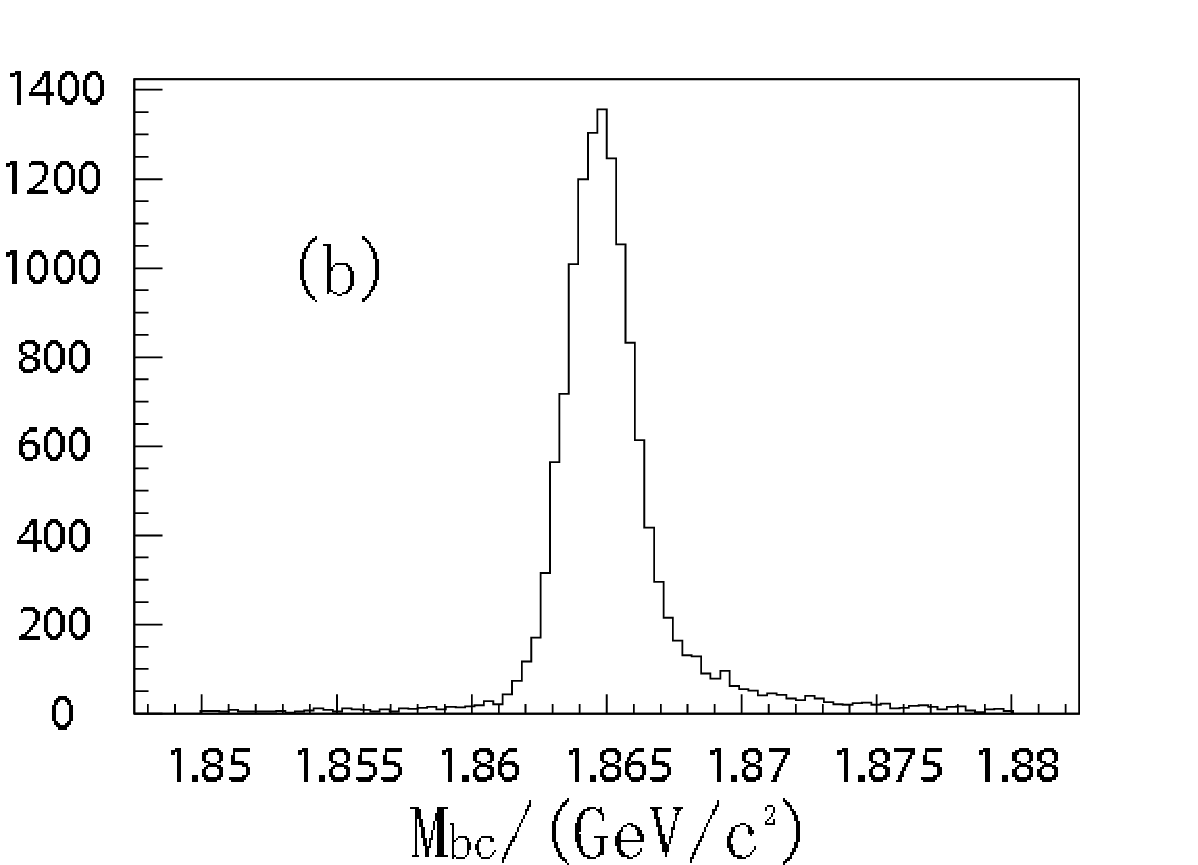}\\

\figcaption{The $M_{bc}$ of (a)the $CP-$ tags; (b)the $CP+$ tags.}
\label{fig:mass CPm}
\end{center}

\subsection{The reconstruction of semileptonic tags}

For tagging the semileptonic decays, we use the decay mode
$D^0\rightarrow K^{-}e^{+}\upsilon_{e}$. To reconstruct the neutral
$D$ meson, good tracks for one electron and one kaon candidate are
required. The good track selection criteria are the same as the $CP$
tagging discussed in Section 3.1. The electron and kaon candidates
are also required to be from the IP, and the likelihood ratio of the
electron and kaon must both be greater than 0.5. Moreover, the two
tracks need to pass a common vertex constraint. After the electron
and pion selections, a standard partial reconstruction technique is
applied to this semileptonic decay channel with one neutrino
associated. Here, we use the ``missing mass" ($U_{miss}$) of
neutrino to select the signal candidates. The ``missing mass" is
defined as follows:
\begin{equation} \displaystyle
U_{miss} \equiv E_{miss} - P_{miss},
\end{equation}
where $E_{miss}=E_{D_{tag}}-E_K - E_{e}$ and
$P_{miss}=|\bm{p}_{D_{tag}}-\bm{p}_{K}-\bm{p}_{e}|$ are the missing
energy and momentum of the neutrino. Here, $E_{K,e}$ and
$\bm{p}_{K,e}$ are the measured energy and momentum of the selected
kaon and electron track. $E_{D_{tag}}$ is the energy of the $D$
meson, which is equal to the beam energy. $\bm{p}_{D_{tag}} = -
\bm{p}_{D_{CP}}$ is the 3-momentum of the $D$ meson, which can be
obtained from the reconstructed momentum of the CP tagged $D$ meson.
For neutrino, the energy and momentum are equal. Thus, the
distribution of $U_{miss}$ must have a mean value at zero. The
distribution of $U_{miss}$ is shown in Fig.~\ref{fig:E-P}. We apply
a $3\sigma$ cut on the $U_{miss}$ to select the semileptonic $D$
decays.
\begin{center}
\includegraphics[width=3.8cm,height=2.8cm]{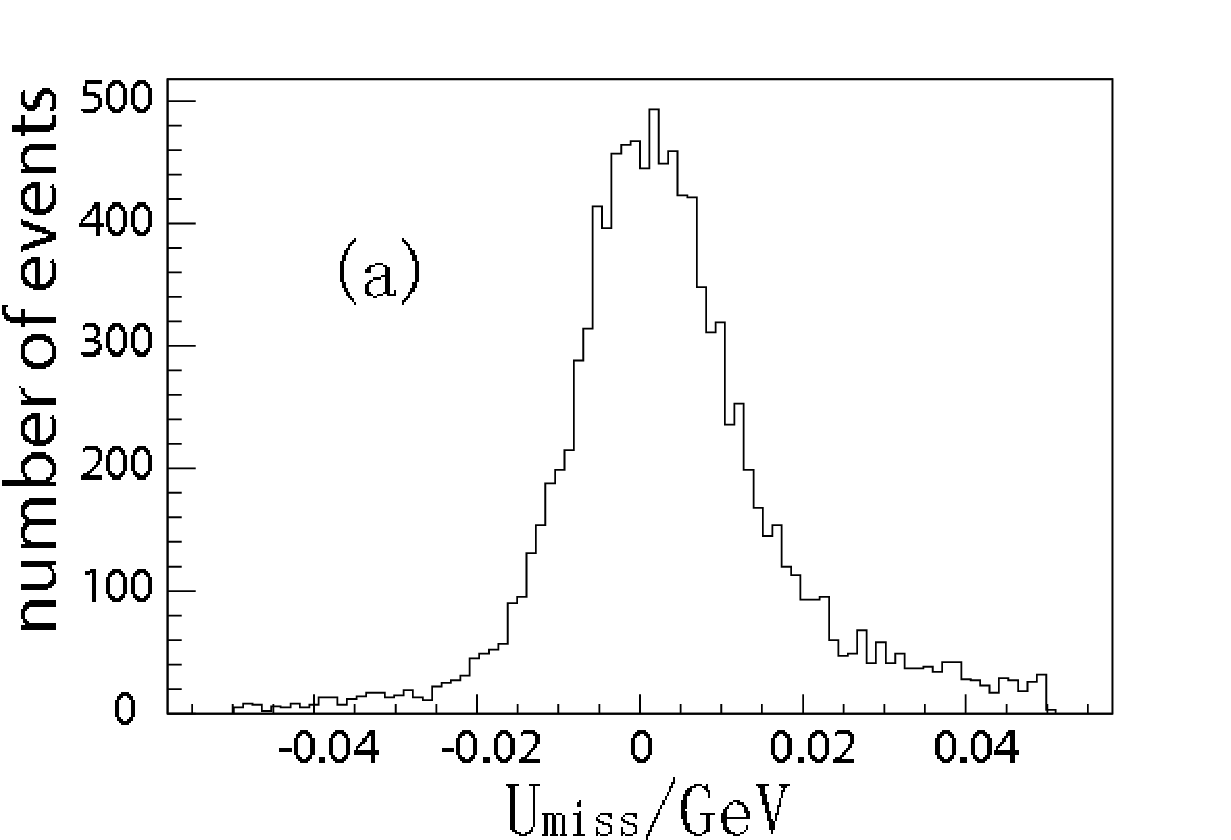}\quad
\includegraphics[width=3.8cm,height=2.8cm]{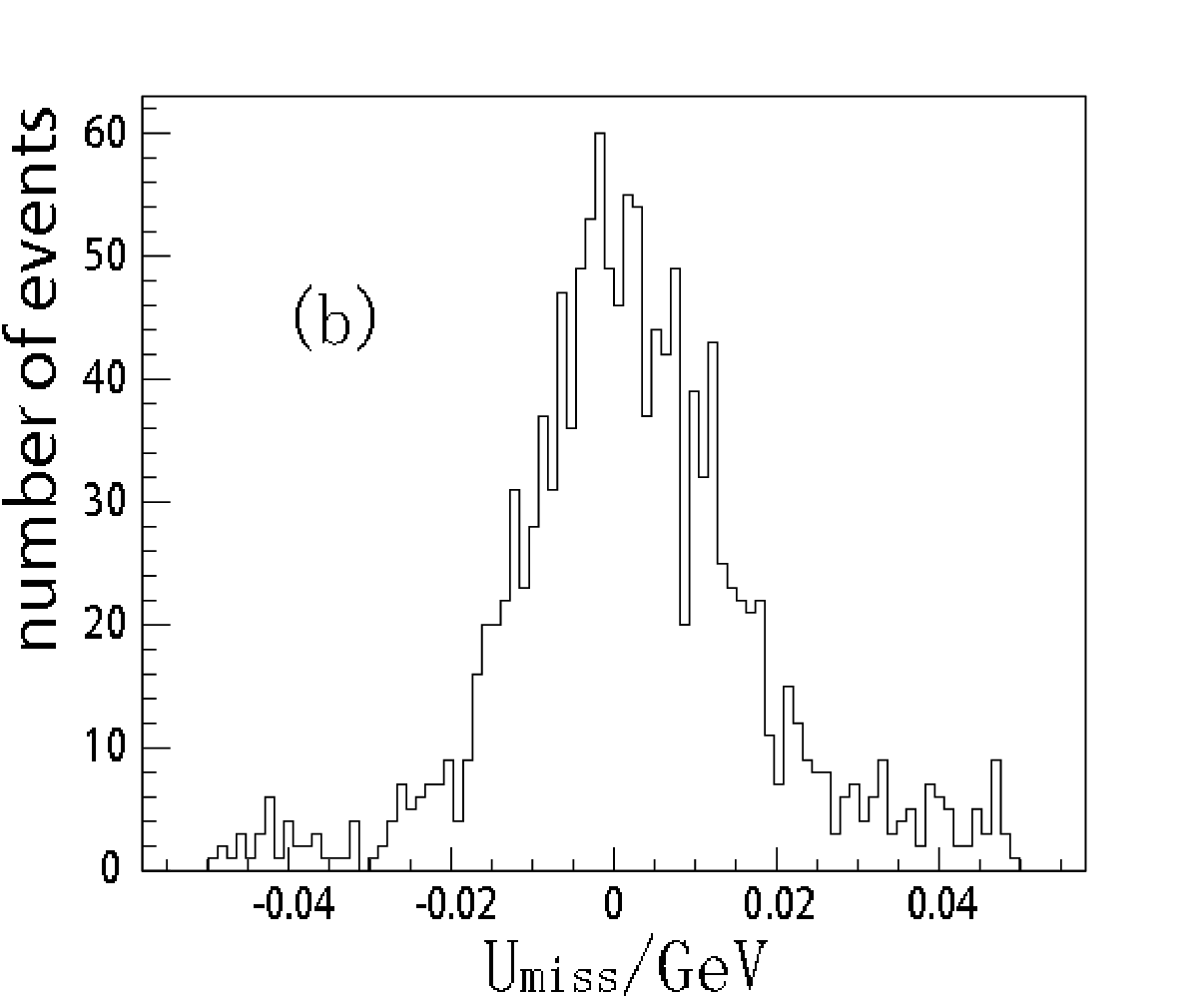}\\
\figcaption{$U_{miss}$ of $\nu_{e}$ for (a) $CP+$ tags and (b) $CP-$
tags.} \label{fig:E-P}
\end{center}

\section{The sensitivity of $y$} \label{sec:sense}
Table \ref{tab:tab_prob} shows the reconstruction efficiency and the
number of estimated doubly-tagged events for different simulated
decay channels. For $\sim20\textrm{fb}^{-1}$ luminosity at
$\psi(3770)$ peak, which approximately corresponds to four years
data taking at BESIII, about $8.0\times 10^{7}$ $\dzdzb$ pairs can
be produced. According to the full simulation, about 11000 doubly
tagged $CP+$ decays and 9000 doubly tagged $CP-$ decays can be
reconstructed.
\begin{center}
\tabcaption{The efficiency and expected events for $8.0 \times
10^{7}$ $\dzdzb$ decays for different decay channels.}
\footnotesize
\label{tab:tab_prob}
\begin{tabular}{|c|c|c|} \hline
decay mode &  efficiency   & event estimation     \\\hline
$K^{-}e^{+}\upsilon_{e}$  & &
\\$K^{-}K^{+},\pi^{-}\pi^{+}$\ &40\% &11701
\\
\hline $K^{-}e^{+}\upsilon_{e}$&  & \\ $K_s\pi^0$  &16.8\%  &7345   \\
\hline
 $K^{-}e^{+}\upsilon_{e}$& & \\$K_s\eta$  &7.7\%  &715   \\ \hline
 $K^{-}e^{+}\upsilon_{e}$& & \\$K_s\eta^\prime$  &4.7\%  &953  \\
\hline
\end{tabular}
\end{center}
For a small y, to calculate the $\sigma_{y}$ of Equation(6), we
ignore the statistical error from single tagged events. Hence, the
statistical error of $y$ parameter can be obtained from the
following equation:
\begin{equation} \displaystyle
\sigma_{y}=\frac{1}{2} \times \sqrt{\frac{1}{N_1}+\frac{1}{N_2}},
\end{equation}
where $N_1$ and $N_2$ represent the reconstructed doubly-tagged
$CP+$ and $CP-$ events. As a result, the $\sigma_y$ is estimated to
be $0.007$ with $\sim20\textrm{fb}^{-1}$ data at $\psi(3770)$ peak
in this analysis. Since the double tagging technique is adopted
here, the background effect can be ignored comparing to the
statistical sensitivity estimated above.

\section{Summary} \label{sec:summary}
In this paper, we presented a MC study on measuring the $\dzdzb$
mixing parameter $y$ at the BESIII experiment. Based on a
$20fb^{-1}$ fully simulated MC sample of $\psi(3770)$ resonance
decays, we estimated the sensitivity of the $y$ measurement to be
0.007. In this analysis, the double tagging technique was used for
reconstructing the $D$ meson pairs. Here, the signal is
reconstructed such that one $D$ decays to $CP$ eigenstates and the
other $D$ decays semileptonically. The electron identification is
essential for this analysis. We improved the e-ID technique for
BESIII experiment, which can also be applied to many other important
physics topics. Our next step is to include more semileptonic decay
modes, such as $D^0\rightarrow K^* e \nu_e$, into this analysis to
improve the sensitivity of y measurement.

\end{multicols}

\vspace{-2mm}
\centerline{\rule{80mm}{0.1pt}}
\vspace{2mm}

\begin{multicols}{2}

\end{multicols}

\clearpage

\end{document}